\documentclass{INTERSPEECH2023}

\usepackage{subfigure}
\usepackage{lipsum}
\usepackage{amssymb}
\usepackage{booktabs}
\usepackage{color}
\usepackage{csquotes}
\interspeechcameraready

\title{EmotionNAS: Two-stream Neural Architecture Search \\ for Speech Emotion Recognition}
\name{Haiyang Sun$^{1,2*}$, Zheng Lian$^{2*}$, Bin Liu$^{2^\dag}$, Ying Li$^2$, Licai Sun$^{1,2}$, Cong Cai$^{1,2}$, Jianhua Tao$^{3^\dag}$, Meng Wang$^4$, Yuan Cheng$^4$}
\address{
 $^1$School of Artificial Intelligence, University of Chinese Academy of Sciences\\
 $^2$State Key Laboratory of Multimodal Artificial Intelligence Systems\\
 $^3$Department of Automation, Tsinghua University, $^4$Ant Group
}
\email{\{sunhaiyang2021, lianzheng2016\}@ia.ac.cn}

\begin{document}
\maketitle
 \newcommand\blfootnote[1]{%
\begingroup
\renewcommand\thefootnote{}\footnote{#1}%
\addtocounter{footnote}{-1}%
\endgroup
}
\blfootnote{*Equal contribution  $^\dag$Corresponding Author}
\begin{abstract}
Speech emotion recognition (SER) is an important research topic in human-computer interaction. Existing works mainly rely on human expertise to design models. Despite their success, different datasets often require distinct structures and hyperparameters. Searching for an optimal model for each dataset is time-consuming and labor-intensive. To address this problem, we propose a two-stream neural architecture search (NAS) based framework, called \enquote{EmotionNAS}. Specifically, we take two-stream features (i.e., handcrafted and deep features) as the inputs, followed by NAS to search for the optimal structure for each stream. Furthermore, we incorporate complementary information in different streams through an efficient information supplement module. Experimental results demonstrate that our method outperforms existing manually-designed and NAS-based models, setting the new state-of-the-art record.
\end{abstract}

\section{Introduction}
\label{sec:intro}
Speech emotion recognition (SER) has received increasing attention due to its contribution to human-computer interactions \cite{neumann2019improving, huang2018speech, lian2018speech}. SER aims to understand how humans express their emotions and then classify each utterance into its emotional state \cite{el2011survey, khalil2019speech}. Existing works are mainly manually-designed models \cite{ma2018emotion, han2018towards, wu2019speech, zhao2021combining}. Despite their success, these works rely on historical experience to design model structures, which is often time-consuming and labor-intensive \cite{xu2019pc, pham2018efficient}. Therefore, how to design networks more intelligently has been brought into focus. To this end, we explore neural architecture search (NAS) for SER \cite{elsken2019neural, chen2019progressive}. By setting the search space, the search strategy, and the evaluation metric, we can optimize the model architecture automatically with little human intervention.

Previously, search methods based on reinforcement learning \cite{zoph2017neural} or evolutionary algorithms \cite{real2017large} avoided the exhaustive search and provided deep learning-inspired design principles. However, these methods need the retraining process every time a new substructure is sampled, which consumes numerous computational resources. To speed up the search process, researchers propose differentiable architecture search (DARTS) \cite{xu2019pc, liu2018darts, Gdarts, Pdarts}, which relaxes the search space to be continuous by applying a softmax operation. Additionally, some efforts \cite{NASbench101, klyuchnikov2020bench} build a model library by pre-training multiple structures. These methods successfully alleviate the time-consuming process in previous works. In this paper, we rely on more efficient search methods to design SER models.

In addition to model design, another challenge in SER is how to represent audios \cite{vinola2015survey}. Among existing handcrafted features, spectrograms are widely utilized in SER \cite{khorram2017capturing, li2018attention, Coattention_22ICASSP, song2022multi}. However, the spectrogram loses some phase information during calculation. Recently, the deep feature, wav2vec \cite{schneider2019wav2vec}, has demonstrated its effectiveness in speech representation learning \cite{siriwardhana2020multimodal}. It is a self-supervised framework that can learn powerful acoustic representations with the help of large amounts of unlabeled data. However, wav2vec may lose some emotion-related information due to different training objectives. To obtain a more comprehensive speech representation, we integrate handcrafted and deep features via a two-stream framework.

Meanwhile, how to fuse multiple features also affects the classification performance. Unsuitable fusion approaches cannot effectively utilize the complementary information in different features, which may lead to performance degradation compared to the best-performing feature (denoted as the dominant feature in this paper). Therefore, we further design an effective fusion method, called ``information supply module (ISM)'', to ensure that the performance of the dominant feature is not degraded and can be further improved.

\begin{figure}[t]
\centering
\includegraphics[width=\linewidth]{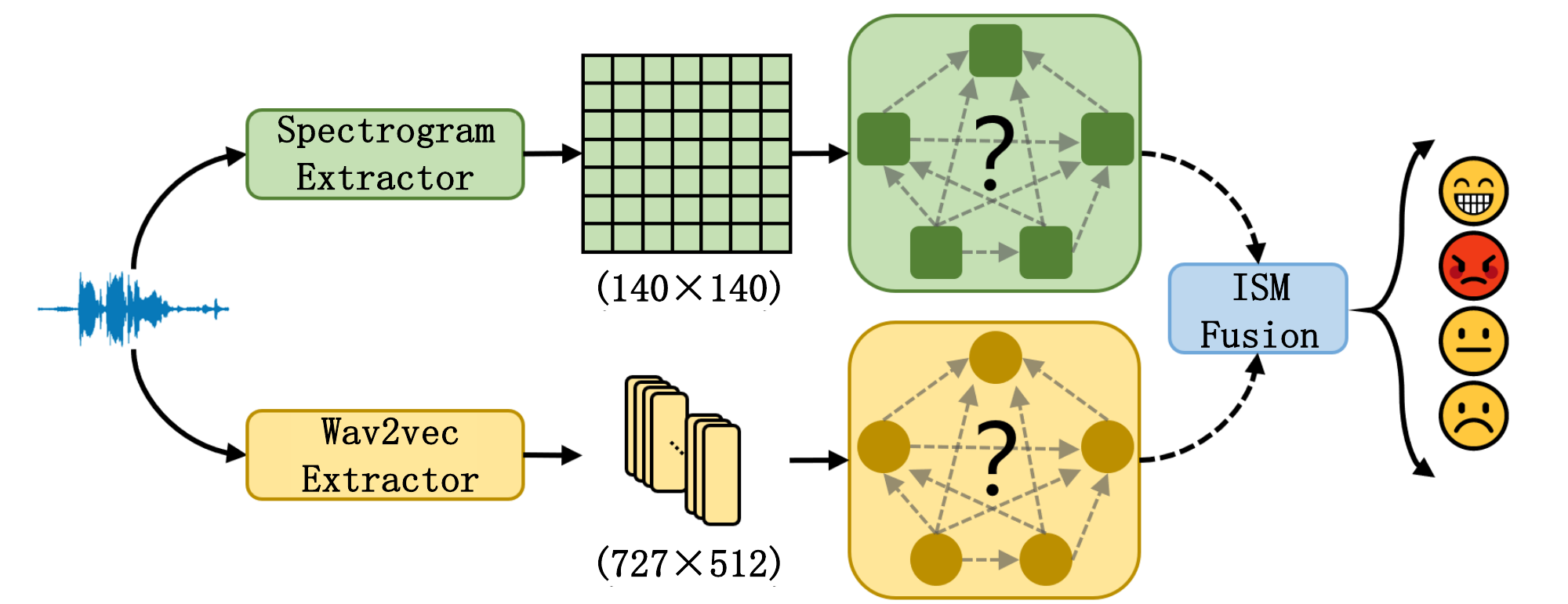}
\caption{The overall structure of EmotionNAS. It is a two-stream architecture that takes spectrogram and wav2vec as the inputs, followed by NAS to design the model automatically. We further fuse the outputs of two branches through ISM to achieve better performance.}
\label{Figure1}
\end{figure}

In summary, to address the low efficiency of the manually-designed approach and inefficient fusion of different features, we propose a novel two-stream framework called \enquote{EmotionNAS}. Figure \ref{Figure1} shows the overall structure of our proposed method. Experimental results show that EmotionNAS outperforms existing manually-designed and NAS-based models. The contributions of this paper can be summarized as follows:
\begin{itemize}
\item We propose a novel framework called \enquote{EmotionNAS}. It incorporates complementary information in handcrafted and deep features, followed by NAS to efficiently search through numerous possible networks to find the optimal structure.

\item We further design ISM to effectively fuse complementary information in different features.

\item Experimental results on IEMOCAP show that our method successfully outperforms existing manually-designed and NAS-based models, setting the new state-of-the-art record.
\end{itemize}

\section{Methodology}
\label{sec2}
EmotionNAS is a two-stream framework that combines handcrafted and deep features for SER. To search for the optimal structure, we explore NAS in model design. Subsequently, we employ ISM to fuse two-stream features effectively.

\subsection{Architecture Search for Spectrogram}
To better utilize the time-frequency two-dimensional information in the spectrogram, our method automatically optimizes the CNN-based structure through an efficient architecture search algorithm DARTS \cite{liu2018darts}. It divides the whole network into normal and reduction cells. All operations adjacent to the reduction cells have a stride of 2, while others have a stride of 1.

\begin{figure}[t]
	\begin{center}
		\subfigure[]{
			\label{Figure2_a}
			\centering
			\includegraphics[width=0.28\linewidth]{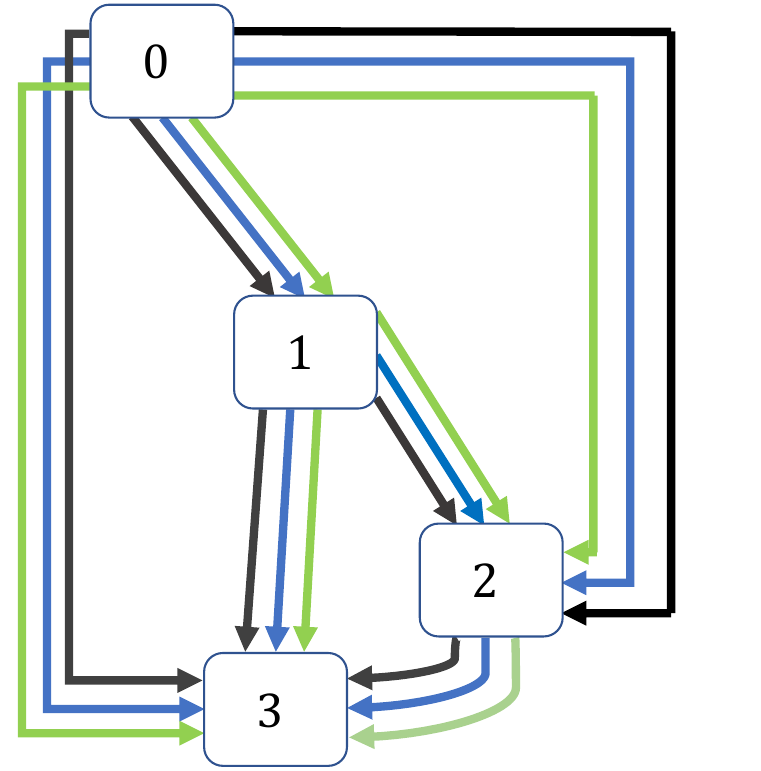}
		}
		\subfigure[]{
			\label{Figure2_b}
			\centering
			\includegraphics[width=0.28\linewidth]{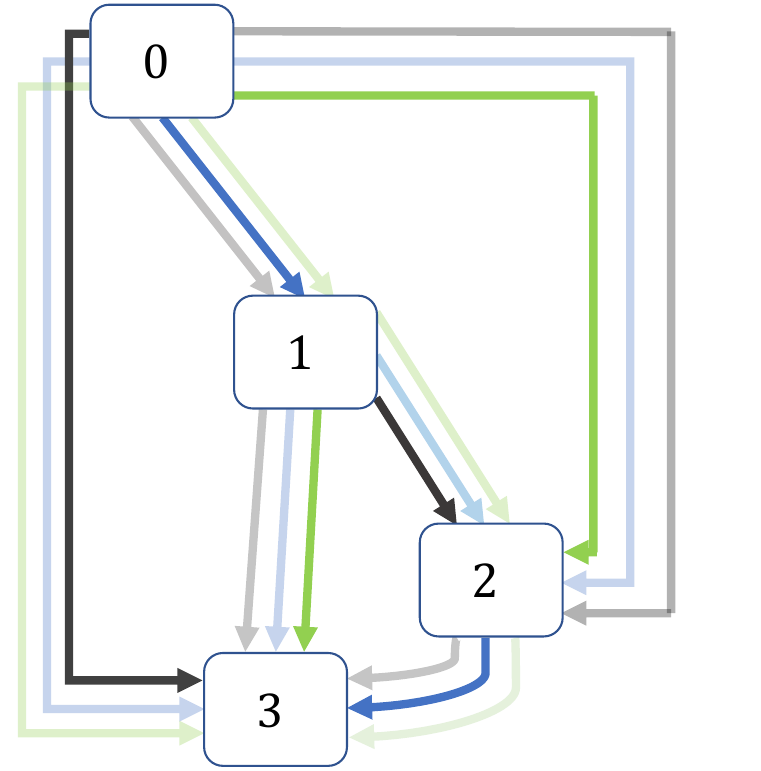}
		}
		\subfigure[]{
			\label{Figure2_c}
			\centering
			\includegraphics[width=0.28\linewidth]{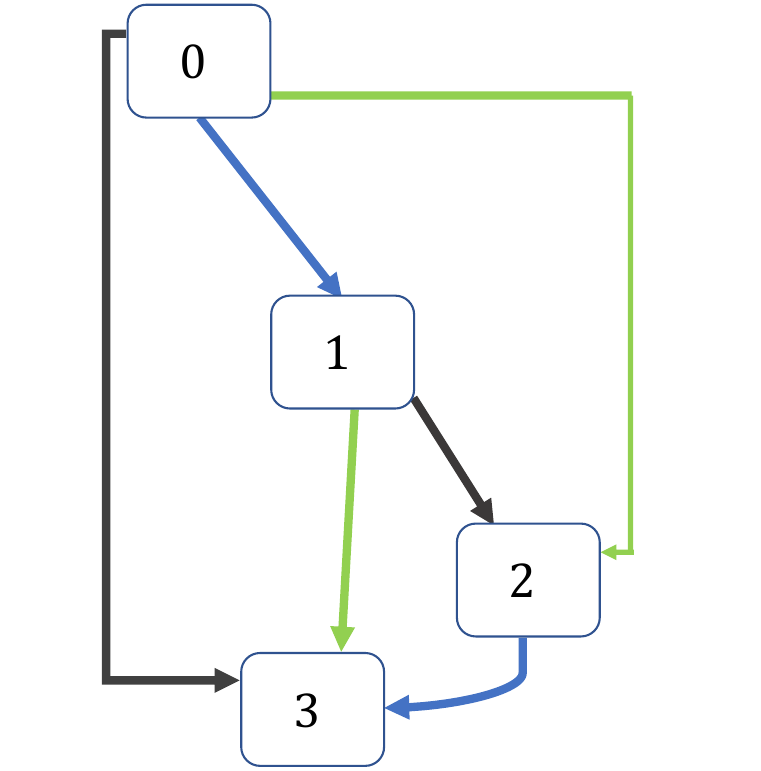}
		}
	\end{center}
	\vspace{-0.5cm}
	\caption{DARTS architecture search in a cell: (a) the initial state of a cell with an ordered sequence of $N=4$ nodes and $\left| \mathbb{O} \right|=3$ operations. Different-colored arrows represent distinct operations; (b) during training, some operations' weights are reduced, resulting in lighter-colored lines; (c) only the operations with the highest weight are retained.}
	\label{Figure2}
\end{figure}

Figure \ref{Figure2} shows the overall search process in a cell. Suppose each cell consists of an ordered sequence of $N$ nodes. Let $x^{(i)}$ denotes the latent feature of the $i$-{th} node. Assume $\mathbb{O}$ is a set of candidate operations, and $\left| \mathbb{O} \right|$ represents the number of operations. To make the search space continuous, DARTS incorporates all possible operations through weights. We take the operations from $x^{(i)}$ to $x^{(j)}$ as an example:
\begin{align}
\alpha_o^{(i,j)} & = \mathrm{softmax}\left(\theta_o^{(i,j)}\right),\\
x^{(j)} & = {
\sum_{o \in \mathbb{O}}{
\alpha_o^{(i,j)}
}o(x^{(i)})
},
\end{align}
where $\theta_o^{(i,j)}$ and $\alpha_o^{(i,j)}$ represent the raw and normalized weights of the operation $o \in \mathbb{O}$ from $x^{(i)}$ to $x^{(j)}$, respectively. The task of architecture search reduces to learning a set of continuous weights $\{ \alpha_o^{(i,j)} \}$. At the end of the search, we only keep the operation with the highest weight.

\subsection{Architecture Search for Wav2vec}
Since wav2vec contains temporal information, we search for the optimal RNN-based structure for this branch \cite{klyuchnikov2020bench}. The search space includes all regular operations in RNN, such as linear mapping, blending, activation function, and element-wise operation. The optimized cell structure is shared across different timesteps. For timestep $t$, the initial nodes consist of the input vector $x_t$ and two hidden states $h_{t-1}^1$ and $h_{t-1}^2$. We aim to optimize the cell architecture and generate new hidden states for the next timestep, i.e., $h_{t}^1$ and $h_{t}^2$ (see Section \ref{sec:4-6} for details).

At the end of the search, we feed wav2vec into the optimized recurrent neural architecture to generate frame-level representations. Since different frames play different parts in SER, we further exploit an attention mechanism to prioritize important frames and fuse them \cite{lian2021ctnet}.

\subsection{Information Supplement Module}
\label{ISM config}
After finding the optimal structure for each stream, we propose an ISM to fuse complementary information from two streams. To ensure that the performance of the dominant feature is not degraded after fusion, we abandon to compress this feature. Instead, we select complementary information from other inputs and fuse them with the dominant feature.

Specifically, suppose the outputs of two branches are denoted as $X_1 \in \mathbb{R}^{d_1}$ and $X_2 \in \mathbb{R}^{d_2}$, respectively. Here, $d_1$ and $d_2$ are the feature dimensions of $X_1$ and $X_2$. Suppose $X_2$ is the dominant feature. The calculation formula of our ISM can be summarized as follows:
\begin{equation}
H_1 = \mbox{Proj}\left(X_1\right),
\end{equation}
\begin{equation}
A_1 = \tanh \left(\mbox{Proj}\left(X_2\right) \odot H_1\right),
\end{equation}
\begin{equation}
F = X_2 + A_1 \odot H_1,
\end{equation}
where $\mbox{Proj}(\cdot)$ and $\odot$ represent a linear transformation function and an element-wise multiplication, respectively. Here, $H_1 \in \mathbb{R}^{d_2}$ and $A_1 \in \mathbb{R}^{d_2}$. In the end, we leverage the fused feature $F \in \mathbb{R}^{d_2}$ for emotion recognition.

\section{Experimental Database and Setup}
\label{sec4}
We first describe the benchmark dataset in our experiments. Following that, we illustrate the feature extraction process, implementation details, and various current advanced baselines.

\subsection{Database}
IEMOCAP \cite{busso2008iemocap} is a benchmark dataset for SER. It consists of five sessions, each with two speakers. For a fair comparison, we adopt five-fold cross-validation with the leave-one-session-out strategy \cite{zhao2021combining}. Eight speakers from four sessions are used as the training set. One speaker in the remaining session is used as the validation set and the other as the test set. We evaluate four emotions (i.e., \emph{neutral}, \emph{angry}, \emph{happy} and \emph{sad}) on improvised data, in line with previous works \cite{zhao2021combining}. Due to the imbalanced class distribution, we use unweighted accuracy (UA) as the primary metric and also report weighted accuracy (WA).

\subsection{Feature Extraction}
In EmotionNAS, we take spectrogram and wav2vec as inputs. The extraction processes are described as follows:

\textbf{Spectrogram:} We utilize the librosa toolkit \cite{mcfee2015librosa} to extract spectrograms from audio. We unify the duration of audio into 8 seconds by zero padding and truncation. Then, a spectrogram is extracted with 25ms Hamming windows and 14ms overlap. Finally, we downsample the spectrogram to $(140 \times 140)$ using an average pooling operation.

\textbf{Wav2vec:} We use the pretrained \emph{wav2vec-large} \cite{schneider2019wav2vec} as the acoustic feature extractor. We unify the wav2vec feature to the maximum length by zero padding, followed by average pooling to downsample it to $(727 \times 512)$.

\subsection{Implementation Details}
Our method consists of three key modules: 1) NAS for spectrogram; 2) NAS for wav2vec; 3) ISM for fusion. Since wav2vec can achieve better performance in SER, we treat this feature as the dominant feature. For the search process, we set hyperparameters based on validation performance.

\textbf{NAS for spectrogram:} In each cell, we set the number of nodes $N=4$ and the number of operations $\left| \mathbb{O} \right|=8$. The search space contains regular operations in CNNs, such as $3\times3$ max pooling, $3\times3$ average pooling, skip connection, $3\times3$ separable convolution, $5\times5$ separable convolution, $3\times3$ dilated convolution, $5\times5$ dilated convolution, and no connection. At the same time, we set the initial number of channels $C=6$ and the number of layers $L=3$.

\textbf{NAS for wav2vec:} The search space contains regular operations in RNNs, such as linear function, blending, element-wise product, element-wise sum, tanh function, sigmoid function, and Leaky ReLU function. Previous work \cite{klyuchnikov2020bench} provided various RNN-based cell structures searched on the Penn Tree Bank dataset \cite{marcus1993building}. We choose the structure with the lowest validation loss in the emotion dataset. Meanwhile, we set the number of layers $L=1$ and the hidden feature dimension $H=256$.

\subsection{Baselines}
To verify the effectiveness of our method, we compare the performance of EmotionNAS with various baselines: \textbf{CNN-GRU} \cite{ma2018emotion} combines CNN and GRU to recognize emotional states. \textbf{CTC-RNN} \cite{han2018towards} exploits connectionist temporal classification \cite{graves2006connectionist} to annotate labels for audio segments automatically. \textbf{SeqCap} \cite{wu2019speech} uses the capsule network and GRUs to capture spatial-temporal information. \textbf{PCNSE-SADRN-CTC} \cite{zhao2021combining} models long-range dependencies by combining parallel convolutional layers, squeeze-and-excitation networks, and self-attention dilated residual networks. \textbf{UniformNAS} \cite{22NASSER} employs NAS to search the optimal structure, and uses a uniform path dropout strategy to encourage all structures to be searched equally.

\begin{table}[t]
	\centering
	\caption{Performance of different approaches. Bold font represents the best performance.}
	\label{Table1}
	\begin{tabular}{ccc}
		\hline
		Method & UA(\%) & WA(\%) \\
		\hline
		CNN-GRU \cite{ma2018emotion} 			  & 64.2 	    & 71.5 \\
		CTC-RNN \cite{han2018towards} 			  & 65.7 	    & 64.2 \\
		SeqCap \cite{wu2019speech}                & 59.7        & 72.7 \\
		PCNSE-SADRN-CTC \cite{zhao2021combining}  & 66.3 	    & \textbf{73.1} \\
		UniformNAS \cite{22NASSER}                & 56.9        & 70.5 \\
		EmotionNAS (Ours)						  & \textbf{69.1} 	    & 72.1 \\
		\hline
	\end{tabular}
\end{table}

\begin{table}[t]
	\centering
	\caption{Performance comparison between NAS-based methods and ResNet. $*$ represents our fair comparison by adjusting the model structure with the similar number of parameters.}
	\label{Table2}
	\begin{tabular}{cccc}
		\hline
		\begin{tabular}[c]{@{}c@{}}{Spectrogram}\\ {Branch}\end{tabular} & {Params} & {UA(\%)} & {WA(\%)}  \\
		\hline
		ResNet18      & 11.18M & \textbf{59.4}   & 60.7   \\
		ResNet*       & 1.23M  & 55.2   & 58.4   \\ 
		\hline
		NAS-C4L3      & 0.07M  & 52.1   & 54.4   \\
		NAS-C6L3      & 0.13M  & 57.3   & \textbf{63.2}   \\
		NAS-C8L4      & 0.35M  & 58.6   & 62.2   \\
		\hline
	\end{tabular}
\end{table}

\begin{table}[t]
	\centering
	\caption{Performance comparison between NAS-based method with RNN and LSTM.}
	\label{Table3}
	\begin{tabular}{cccc}
		\hline
		\begin{tabular}[c]{@{}c@{}}{Wav2vec}\\ {Branch}\end{tabular} & {Params} & {UA(\%)} & {WA(\%)}  \\
		\hline
		RNN-H512L2  & 1.05M  & 60.2   & 64.1   \\
		RNN-H512L1  & 0.53M  & 61.0   & 64.4   \\ 
		LSTM-H512L2 & 4.21M  & 64.6   & 65.7   \\
		LSTM-H512L1 & 2.10M  & 64.6   & 67.7   \\ 
		\hline
		NAS-H512L1  & 2.37M  & \textbf{67.2}   & 68.4   \\
		NAS-H256L1  & 0.73M  & 66.2   & \textbf{70.3}   \\ 
		\hline
	\end{tabular}
\end{table}

\begin{table}[t]
	\centering
	\caption{Importance of each branch.}
	\label{Table4}
	\begin{tabular}{ccc}
		\hline
		Method & {UA(\%)} & {WA(\%)}  \\
		\hline
		spectrogram branch  &57.3      &63.2 \\
		wav2vec branch      &66.2      &70.3 \\
		EmotionNAS (Ours)   &\textbf{69.1}  &\textbf{72.1} \\
		\hline
	\end{tabular}
\end{table}

\section{Results and Discussion}
\label{sec5}
In this section, we first conduct comparative experiments with currently advanced systems to verify the effectiveness of our method. Then, we systemically investigate the importance of each module in EmotionNAS, including the NAS-based approach, two-stream framework, and ISM fusion strategy. Finally, we visualize the confusion matrices and search results.

\subsection{Comparison with Existing Works}
To verify the effectiveness of our method, we treat current advanced approaches as our baselines. These methods include existing manually-designed and NAS-based models with consistent data division. We report baseline results according to their original papers. Experimental results in Table \ref{Table1} show that EmotionNAS outperforms most existing methods and achieves a performance improvement of 2.8\% over manually-designed models \cite{zhao2021combining}. Compared with these approaches, our NAS-based framework can optimize structures and find more effective models than manually-designed approaches, automatically.

Meanwhile, we observe that EmotionNAS outperforms UniformNAS \cite{22NASSER}, a NAS-based approach, by 12.2\% on UA and 1.2\% on WA. But limited by handcrafted features, this baseline cannot leverage the useful information from deep features. Our framework, by contrast, not only enables efficient network design but also effectively fuses handcrafted and deep features, bringing better performance on SER.

\subsection{Advantage of NAS}
In this section, we compare the performance of several NAS-based and manually-designed models. Experimental results are shown in Table \ref{Table2}$\sim$\ref{Table3}. For the spectrogram branch, NAS-based methods can achieve better performance with fewer parameters (see Table \ref{Table2}). For the wav2vec branch, our method also outperforms RNN and LSTM in emotion recognition (see Table \ref{Table3}). These results reveal the advantage of NAS in model design.

\subsection{Necessity of Two Branches}
In this section, we further reveal the necessity of two branches. As shown in Table \ref{Table4}, spectrogram and wav2vec achieve 57.3\% and 66.2\% on UA, 63.2\% and 70.3\% on WA, respectively. After multi-branch fusion, we further improve UA to 69.1\% and WA to 72.1\%. These results confirm that handcrafted and deep features contain complementary information in emotion recognition. Through the fusion process, we can achieve better classification performance on SER.

\subsection{Effectiveness of ISM}
To verify the effectiveness of our method, we compare ISM with other fusion strategies. Followed with the symbols in Section \ref{ISM config}, we denote the outputs of two branches as $X_1$ and $X_2$. Here, $X_1$ is the dominant feature and $H_1$ is the linear transformation of $X_1$. \textbf{Concat} concatenates $X_1$ and $X_2$. \textbf{Sum}, \textbf{Max} and \textbf{Min} denote the corresponding operations on $H_1$ and $X_2$. 

Experimental results are listed in Table \ref{Table5}. Among the four comparison strategies, only \textbf{Min} and \textbf{Concat} surpass the performance of the dominant branch. It indicates that most strategies cannot effectively fuse complementary information in different features. But through ISM, we can significantly improve the performance of UA and WA. These results demonstrate the effectiveness of our fusion strategy.


\begin{table}[t]
	\centering
	\caption{Importance of ISM.}
	\label{Table5}
	\begin{tabular}{ccc}
		\hline
		Method & {UA(\%)} & {WA(\%)}  \\
		\hline
		Dominant branch (wav2vec)  &66.2  &70.3 \\
		Concat              &67.1      &68.5 \\
		Sum                 &63.8     &66.0 \\
		Max                 &65.0     &67.5 \\
		Min                 &68.1       &65.2 \\
		EmotionNAS (Ours)   &\textbf{69.1}  &\textbf{72.1} \\
		\hline
	\end{tabular}
\end{table}


\begin{figure}[t]
	\begin{center}
		\subfigure[]{
			\label{Figure4_a}
			\centering
			\includegraphics[width=0.46\linewidth]{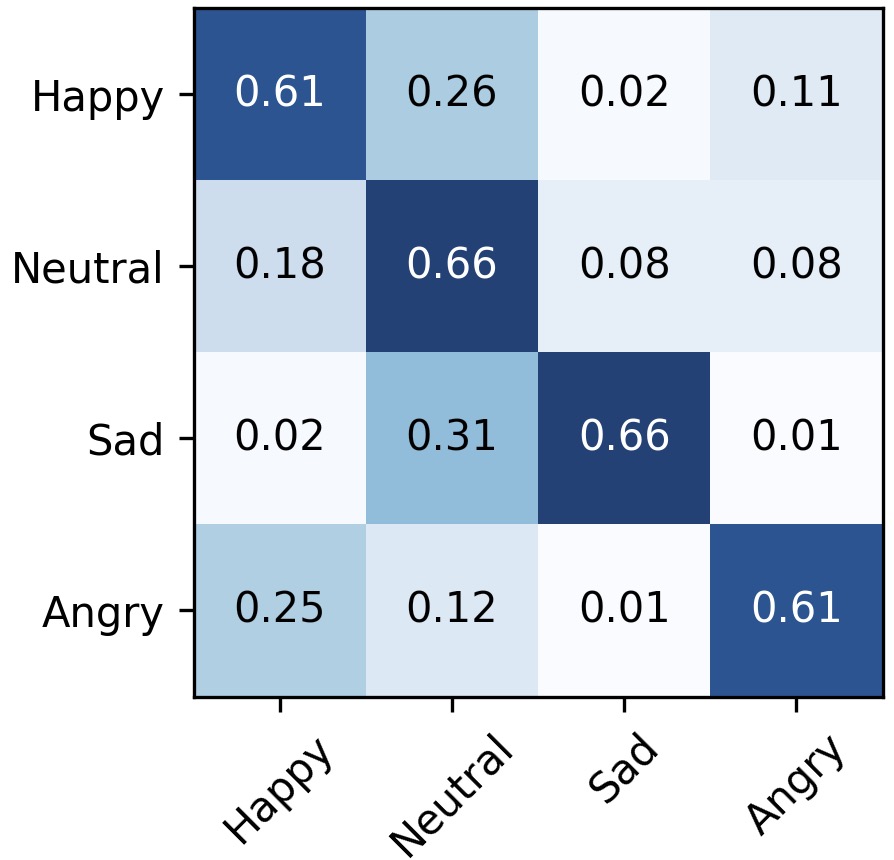}
		}
		\subfigure[]{
			\label{Figure4_b}
			\centering
			\includegraphics[width=0.46\linewidth]{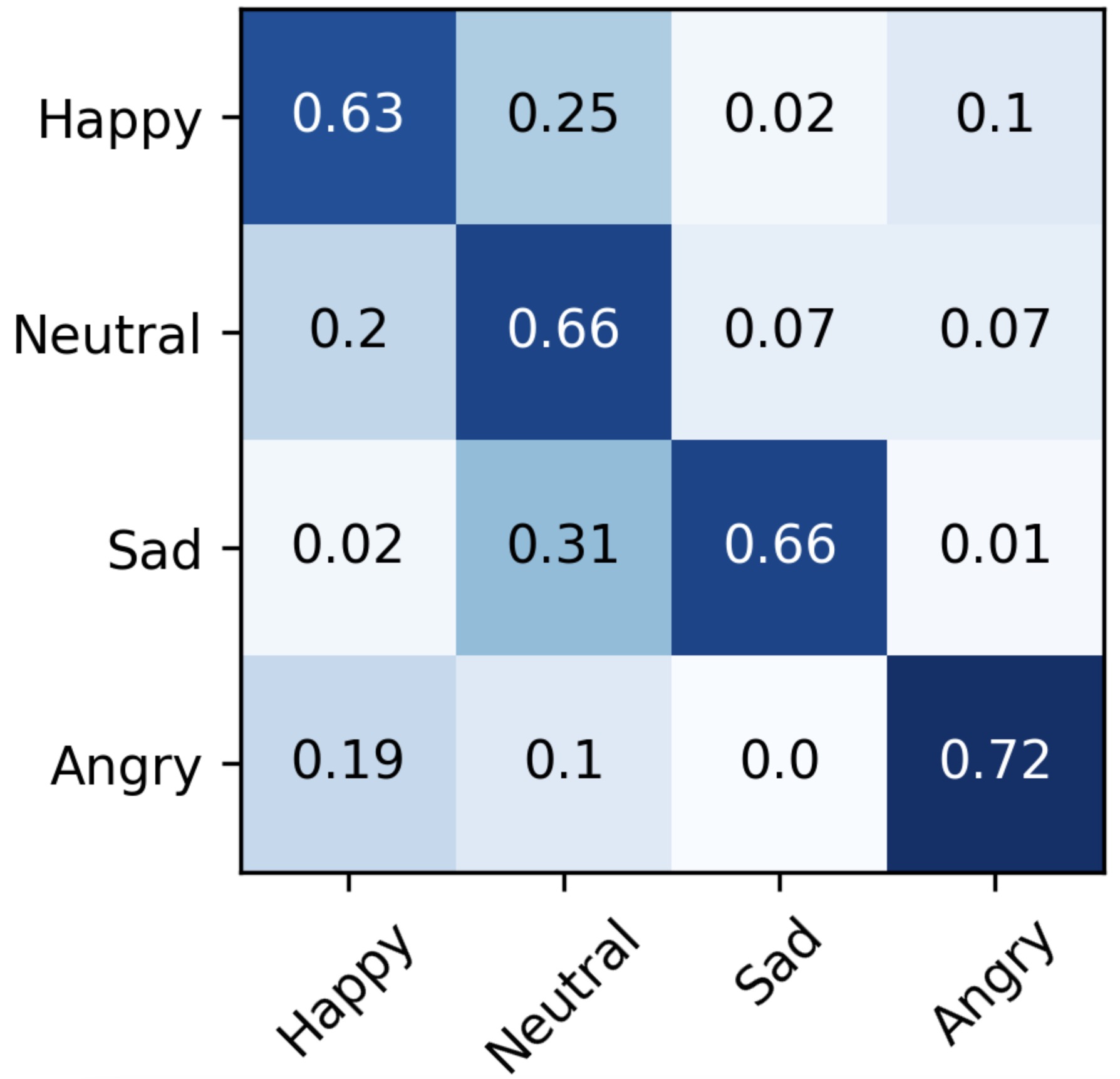}
		}
	\end{center}
	\vspace{-0.5cm}
	\caption{Visualization of confusion matrices. (a) Results of \textbf{Min}. (b) Results of EmotionNAS.}
	\label{Figure4}
\end{figure}

\begin{figure}[!h]
	\begin{center}
		\subfigure[]{
			\label{Figure5_1}
			\centering
			\includegraphics[width=0.92\linewidth]{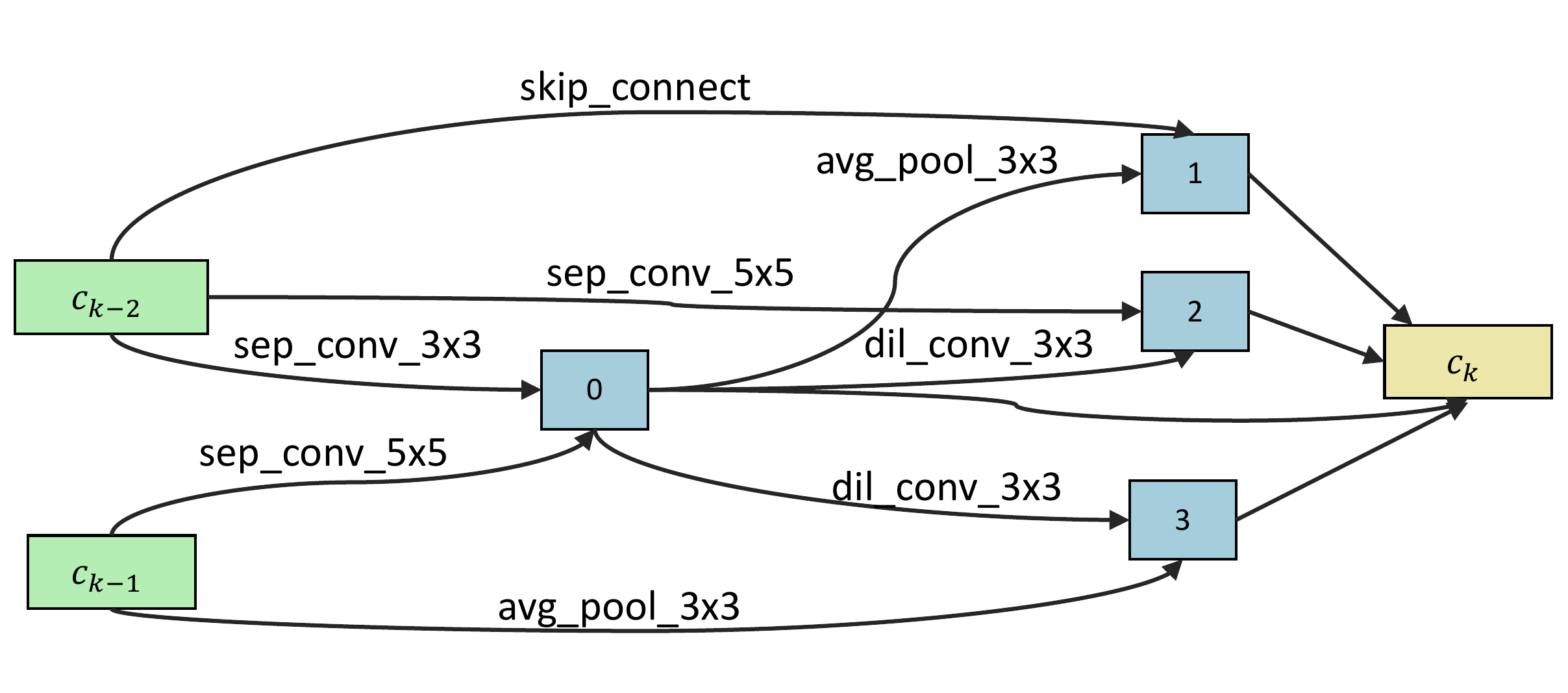}
		}
		\subfigure[]{
			\label{Figure5_2}
			\centering
			\includegraphics[width=0.92\linewidth]{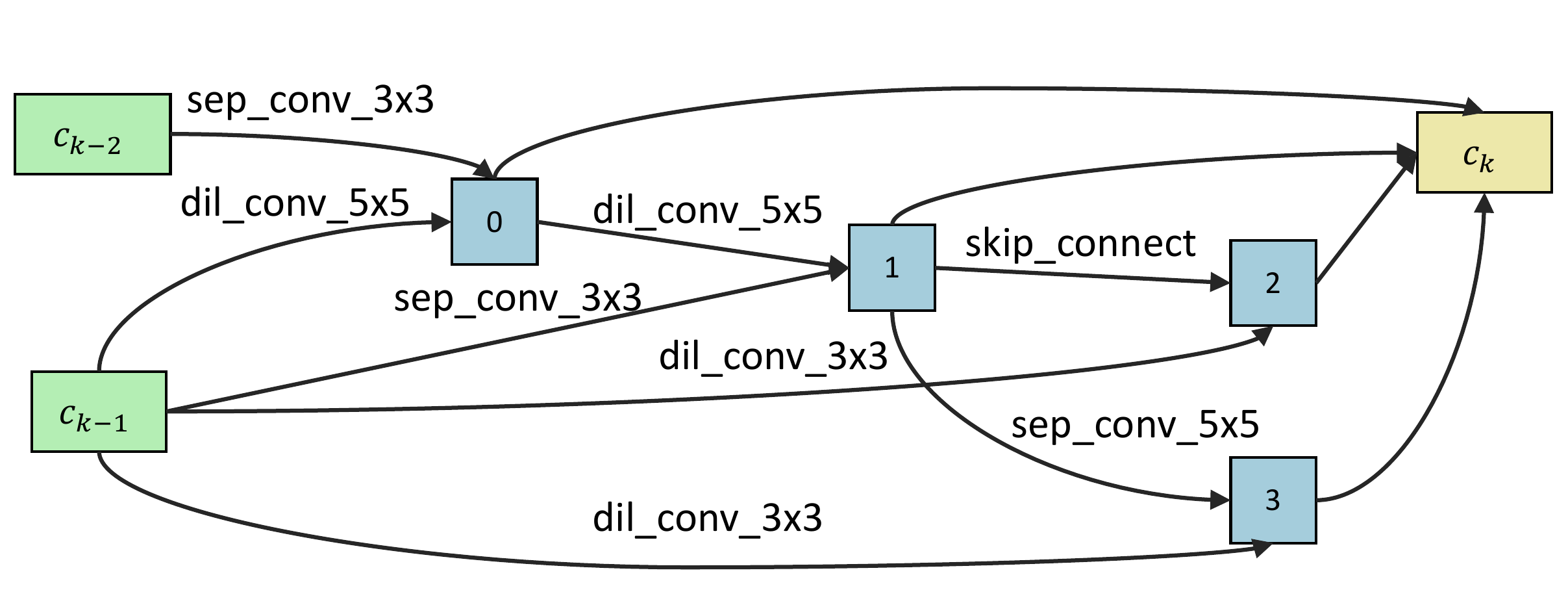}
		}
		\subfigure[]{
			\label{Figure5_3}
			\centering
			\includegraphics[width=0.92\linewidth]{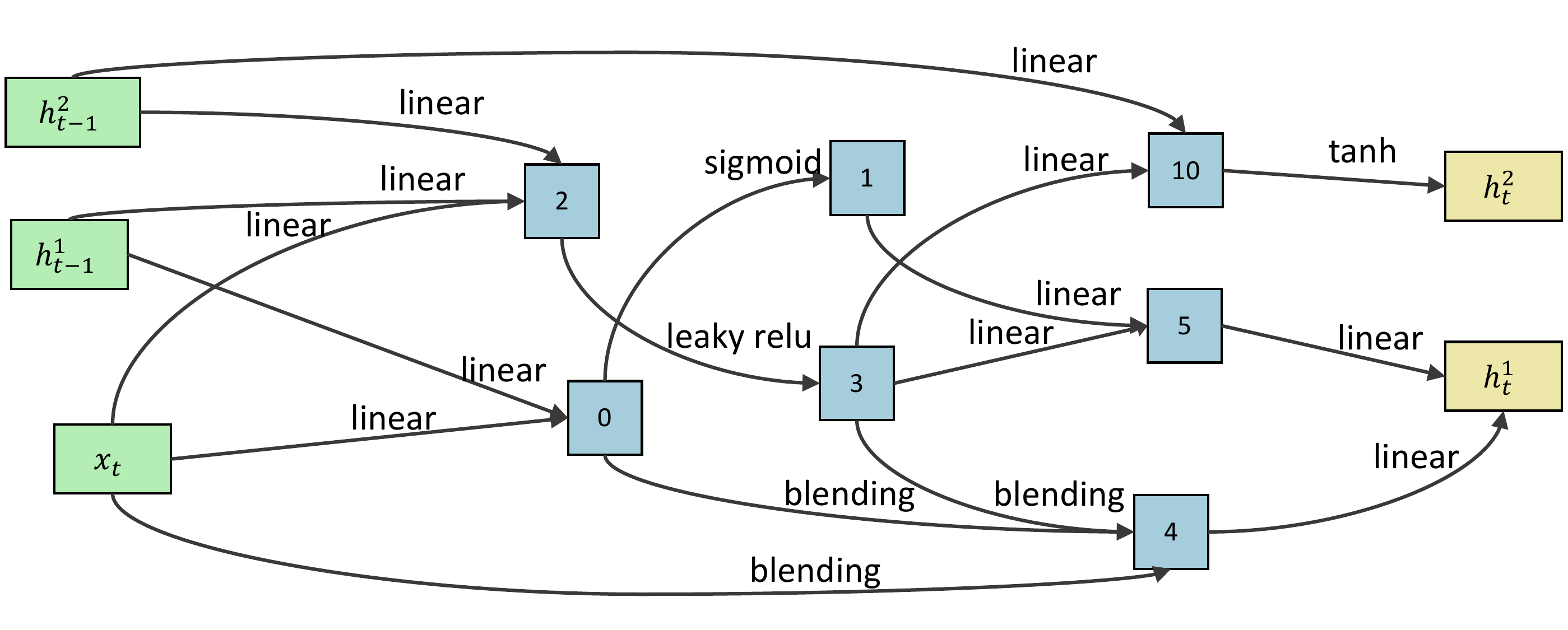}
		}
	\end{center}
	\vspace{-0.5cm}
	\caption{Visualization of search results: (a) and (b) are search results of the normal cell and the reduction cell for the spectrogram branch; (c) is the search result for the wav2vec branch.}
	\label{Figure5}
\end{figure}

\subsection{Visualization of Confusion Matrices}
Figure \ref{Figure4} shows the confusion matrices for \textbf{Min} and EmotionNAS. We observe that EmotionNAS achieves better performance in most emotion categories, especially \emph{happy} and \emph{angry}. Our findings also highlight the limitations of traditional fusion strategies. Compared with \textbf{Min}, our fusion method ISM can effectively fuse multiple features and achieve better performance.

\subsection{Visualization of Search Results}
\label{sec:4-6}
In this section, we further visualize the search results for each branch. Figure \ref{Figure5_1}$\sim$\ref{Figure5_2} show the search results for the spectrogram branch. There are two types of cells: the normal cell and the reduction cell. For cell $k$, the initial nodes consist of the outputs in previous cells, i.e., $c_{k-2}$ and $c_{k-1}$. Figure \ref{Figure5_3} shows the search result for the wav2vec branch. For timestep $t$, the initial nodes consist of the input vector $x_t$ and two hidden states $h_{t-1}^1$ and $h_{t-1}^2$. We aim to generate new hidden states for the next timestep, i.e., $h_{t}^1$ and $h_{t}^2$.

\section{Conclusions}
\label{sec6}
In this paper, we propose a novel framework for SER, called ``EmotionNAS''. It takes spectrogram and wav2vec as the inputs, followed by NAS to optimize the network structure automatically. We further design ISM to integrate complementary information in different branches. Experimental results demonstrate that our method outperforms existing manually-designed and NAS-based methods. Meanwhile, we also systemically prove the importance of each component in EmotionNAS, including the NAS-based method, the two-stream framework and the ISM fusion strategy.

\section{Acknowledgments}

This work is supported by the National Natural Science Foundation of China (NSFC) (No.61831022, No.62276259, No.62201572, No.U21B2010), Beijing Municipal Science\&Technology Commission, Administrative Commission of Zhongguancun Science Park No.Z211100004821013, Open Research Projects of Zhejiang Lab (NO. 2021KH0AB06), CCF-Baidu Open Fund (No.OF2022025).

\bibliographystyle{IEEEtran}
\bibliography{mybib}
\end{document}